\documentclass[12pt]{article}
\usepackage{latexsym,amsmath,amsfonts,amssymb}
\newtheorem{Theorem}{Theorem}
\def\e{\epsilon}
\def\p{\partial}
\author{Renat Zhdanov\thanks{E-mail:\ renat.zhdanov@bio-key.com}\\
BIO-key International, Eagan, MN, USA}
\title{On group classification of evolution equations admitting
non-local symmetries }
\date{}

\begin{document}
\maketitle

\begin{abstract}
We prove that any evolution equation admitting a potential symmetry
can always be reduced to another evolution equation such that the
potential symmetry in question maps into the group of its contact
symmetries. Based on this fact is out group approach to
classification of evolution equations possessing non-local
symmetries. We present several examples of classifications of
second-order evolution equations admitting potential symmetries.
\end{abstract}

\section{Introductory Remarks}

We study evolution type equations in one spatial dimension
\begin{equation}
\label{1}
u_t=F(t,x,u,u_1,u_2,\ldots,u_n),\quad n\ge2,
\end{equation}
where $u=u(t,x)$ is a real-valued function of two real variables $t,
x$, $u_i={\partial^i u}/{\partial x^i},\ i=1,2,\ldots, n$, and $F$
is an arbitrary smooth real-valued function.

We say that partial differential equation (\ref{1}) admits a Lie
symmetry if there is a one-parameter family of maps ${\mathbb
R}^3\to{\mathbb R}^3$
\begin{equation}
\label{2} t'=T(t,x,u,\theta),\quad x'=X(t,x,u,\theta),\quad
u=U(t,x,u,\theta),
\end{equation}
that transform the solution set of equation (\ref{1}) into itself.
One of the fundamental results of Sophus Lie are his celebrated
theorems revealing group nature of such maps and establishing the
infinitesimal criterium for any partial differential equations to
admit a Lie symmetry see (e.g., \cite{ovs1}-\cite{blu1}).

If we allow for the right-hand sides of (\ref{2}) to depend on the
first derivatives, namely,
\begin{eqnarray}
&&t'=T(t,x,u,u_t,u_x,\theta),\quad
x'=X(t,x,u,u_t,u_x,\theta),\nonumber\\
&&u'=U(t,x,u,u_t,u_x,\theta),\label{3}\\
&& u_t'=V(t,x,u,u_t,u_x,\theta),\quad
u_x'=W(t,x,u,u_t,u_x,\theta)\nonumber
\end{eqnarray}
requiring that the first-order tangency condition
\begin{equation}
\label{13}
du-u_tdt-u_xdx=0
\end{equation}
is invariant under the action of (\ref{3}), then we arrive at the
concept of contact transformation group \cite{ibr1}.

Following Lie's methodology we rewrite (\ref{3}) in the
infinitesimal form
\begin{eqnarray}
&&t'=t + \e \tau(t,x,u,u_t,u_x),\quad x'=x +
\e\xi(t,x,u,u_t,u_x),\nonumber\\
&&u'=u+\e \eta(t,x,u,u_t,u_x),\label{4}\\
&& u_t'= u_t + \e \zeta(t,x,u,u_t,u_x),\quad u_x'=u_x + \e
\rho(t,x,u,u_t,u_x).\nonumber
\end{eqnarray}
Here
$$
\tau = \left.\frac{dT}{d\theta}\right|_{\theta=0},\quad
\xi = \left.\frac{dX}{d\theta}\right|_{\theta=0}, \quad
\eta = \left.\frac{dU}{d\theta}\right|_{\theta=0}, \quad
\zeta = \left.\frac{dV}{d\theta}\right|_{\theta=0}, \quad
\rho = \left.\frac{dW}{d\theta}\right|_{\theta=0}
$$and $\e$ is an infinitesimal parameter.

It is a common knowledge that the coefficients $\tau,\ldots,\rho$
are expressed in terms of the generating function
$\varphi=\varphi(t,x,u,u_t,u_x)$ \cite{ibr1}
\begin{eqnarray}
&&\tau = -\frac{\p\varphi}{\p u_t},\quad
\xi = -\frac{\p\varphi}{\p u_x},\nonumber\\
&&\eta = \varphi-u_t\frac{\p\varphi}{\p u_t} -
u_x\frac{\p\varphi}{\p u_x}, \nonumber\\
&&\zeta = \frac{\p\varphi}{\p t} + u_t\frac{\p \varphi}{\p u},\quad
\rho = \frac{\p\varphi}{\p x} + u_x\frac{\p \varphi}{\p u}.
\nonumber
\end{eqnarray}
Magadeev proves \cite{mag1} that the most general form of the
contact transformation group admitted by evolution equation
(\ref{1}) under $n\ge 2$ reads as
$$
t'=T(t,\theta),\quad x'=X(t,x,u,u_x,\theta),\quad
u'=U(t,x,u,u_x,\theta).
$$

Lie symmetries plays the central role in the modern theory of
differential equations. The main reason is that the most successful
mathematical models of the physical, chemical and biological
processes do admit nontrivial Lie symmetries. One can argue that it
is this very property which singles out the Maxwell equations
from the set of hyperbolic type systems of first-order linear
partial differential equations \cite{fus1}. This type of
speculations were the major motivation for the so called symmetry
selection principle. It says, that those equations should be
considered as realistic models of the nature processes which admit
nontrivial Lie symmetries (for further discussion of this matter,
see \cite{fus2}). It is customary to call this procedure group
classification of differential equations.

There is a host of papers devoted to group classification of
different subclasses of the class of equations (\ref{1}) (see, e.g.,
\cite{ren1}-\cite{pol1} and references therein). They provide a very
clear picture of what we can do and what we cannot achieve by
utilizing the classical Lie symmetries. Since Lie symmetries cannot
be answers to all challenges of the modern theory of nonlinear
differential equations, one has always been looking for a way to
generalize the notions of Lie and contact symmetries. A natural
approach is allowing for the right-hand sides of (\ref{4}) to depend
on higher derivatives of $u$, which yields the concept of
Lie-B\"acklund symmetry \cite{ibr1}.

A more radical generalization would be to allow for dependance on
integrals of $u$, which is the way the non-local symmetries arose.
The major problem with non-local symmetries is that unlike the case
of Lie or Lie-B\"acklund symmetries there is no generic algorithm
for computing these. Almost each specific class of partial
differential equations requires special, often unique treatment
\cite{ibr1,fus1}. The most well studied is the case of non-local
symmetries of linear partial differential equations (see
\cite{sha1,fus1} and the references therein).

Much less is known about non-local symmetries of nonlinear partial
differential equations. One of the possible approaches has been
suggested by Bluman \cite{blu2,blu3}. The idea is presenting
evolution equation (\ref{1}) in the form of conservation law
\begin{equation}
\label{6}
\frac{\p}{\p t}\Bigl(f(t,x,u)\Bigr) = \frac{\p}{\p
x}\Bigl(g(t,x,u,u_1,\ldots,u_{n-1})\Bigr),
\end{equation}
where $f,g$ are some smooth real-valued functions. Next, one
introduces the new dependant variable $v=v(t,x)$ and rewrite
(\ref{6}) as the system of two differential equations
\begin{equation}
\label{7}
v_t=g(t,x,u,u_1,\ldots,u_n),\quad v_x=f(t,x,u).
\end{equation}
Suppose now that system (\ref{7}) admits a Lie symmetry
\begin{equation}
\label{8}
\begin{array}{l}
t'=T(t,x,u,v,\theta),\quad x'=X(t,x,u,v,\theta),\\[2mm]
u'=U(t,x,u,v,\theta),\quad v'=V(t,x,u,v,\theta),
\end{array}
\end{equation}
and what is more, at least one of the functions $T, X, U, V$ depends
on the non-local variable $v$. Then (\ref{7}) is the non-local
symmetry of the initial equation (\ref{1}) called also its potential
symmetry.

There is another approach to constructing non-local symmetries which
is to apply a non-local transformation to an equation admitting
non-trivial Lie symmetries. Then some of symmetries of the initial
equation will remain Lie symmetries of the transformed equation,
while the others become non-local ones. They are called quasi-local
symmetries \cite{ibr2,puk}.

Recently, we developed the regular group approach to classification
of systems of evolution equations that admit quasi-local symmetries
\cite{ren2,ren3}. It has been conjectured that there is a link
between quasi-local and potential symmetries. The principal aim of
this paper is to prove that the connection does exist and, in fact,
any potential symmetry is a quasi-local one. Namely, there exists a
non-local change of variables that maps any equation possessing
potential symmetry into an equation, which admits a contact
symmetry. The connection between contact and potential symmetries
provides a convenient tool for group classification of equations
admitting non-local symmetries. We give several examples of such
classification.

\section{Theoretical Background}

Let us introduce new dependent variable $w(t,x)=f(t,x,u(t,x))$ and
rewrite equations (\ref{6}), (\ref{7}) as follows
\begin{equation}
\label{9} v_t=\tilde g(t,x,w,w_1,\ldots,w_n),\quad v_x = w.
\end{equation}
Equation (\ref{1}) now reads as
$$
w_t=\frac{\p}{\p x}\Bigl(\tilde g(t,x,w,w_1,\ldots,w_{n-1})\Bigr)
$$
and its Lie symmetry group (\ref{8}) takes the form
\begin{equation}
\label{10}
\begin{array}{l}
t' = \tilde T(t,x,w,v,\theta),\quad
x' = \tilde X(t,x,w,v,\theta),\\[2mm]
w' = \tilde U(t,x,w,v,\theta),\quad
v' = \tilde V(t,x,w,v,\theta).
\end{array}
\end{equation}
Note, that the above transformation group by definition preserve the
first-order tangency conditions
$$
dv-v_tdt-v_xdx = 0,\quad dw-w_tdt-w_xdx = 0.
$$

Eliminating $w$ from the first equation of system (\ref{9})
yields
\begin{equation}
\label{11}
v_t = \tilde g(t,x,v_1,\ldots,v_n).
\end{equation}
By construction, transformation group (\ref{10}) maps the solution
set of (\ref{11}) into itself. Consequently, this group is a Lie
symmetry of evolution equation (\ref{11}). Moreover, using the
second equation from (\ref{9}) we can eliminate the function $w$
from the relations (\ref{10}) thus getting
\begin{equation}
\label{12}
\begin{array}{l}
t' = \tilde T(t,x,v_x,v,\theta),\quad
x' = \tilde X(t,x,v_x,v,\theta),\\[2mm]
v' = \tilde V(t,x,v_x,v,\theta),\quad
v'_{x'} = \tilde U(t,x,v_x,v,\theta).
\end{array}
\end{equation}
Again, by construction, this group transforms the solution set of
equation (\ref{11}) into itself and therefore is the symmetry group
of (\ref{11}). Next, as group (\ref{10}) preserves the tangency
condition $dv-v_tdt-v_xdx=0$, so does group (\ref{12}). Hence,
transformation group (\ref{12}) is the group of contact
transformations.

Thus we proved the following assertion.
\begin{Theorem}
Let (\ref{1}) be an arbitrary evolution type differential equation
possessing potential symmetry. Then there is a change of variables
that transforms potential symmetry into a contact symmetry group of
the appropriately transformed evolution equation (\ref{11}).
\end{Theorem}
{\bf Note 1.} In the case under consideration the transformed
evolution equation is given by (\ref{11}), while its contact
symmetry group reads as (\ref{12}). The change of variables reducing
(\ref{1}) to (\ref{11}) is given by the formula $v_x=f(t,x,u)$.

\noindent {\bf Note 2.} Transformed evolution equation (\ref{11})
admits Lie symmetry $\p_u$. This important observation will be used
to classify potential symmetries associated with a given evolution
equation.

We established that any potential symmetry of an evolution equation
is quasi-local symmetry in a sense that it can be reduced to a local
(contact) symmetry group. We are going to prove that the inverse
assertion is also true. Suppose that evolution equation (\ref{1})
admits contact symmetry group, which has, at least, a one-parameter
subgroup
\begin{equation}
\label{14}
\begin{array}{l}
t' = t,\quad x' = \tilde X(t,x,u,u_x,\theta),\\[2mm]
u' = \tilde U(t,x,u,u_x,\theta),\quad u'_{x'} = \tilde
U(t,x,u,u_x,\theta)
\end{array}
\end{equation}
preserving the temporal variable $t$. Then there is a contact
transformation \cite{ibr1}
\begin{equation}
\label{19}
\begin{array}{l}
\bar t = t,\quad \bar x = \bar X(t,x,u,u_x,\theta),\\[2mm]
\bar u = \bar U(t,x,u,u_x,\theta),\quad \bar u_{\bar x} = \bar
V(t,x,u,u_x,\theta)
\end{array}
\end{equation}
reducing (\ref{14}) to the one-parameter groups of translations by
$\bar u$. The corresponding evolution equation (\ref{1}) takes the
form
\begin{equation}
\label{15}
u_t = f(t,x,u_1,\ldots,u_n).
\end{equation}
Note that we dropped the bars.

Now following \cite{ren3,ren4} (see, also \cite{sok1}) we
differentiate (\ref{15}) with respect to $x$ and introduce the new
dependent variable $v(t,x)=\p u/\p x$ thus getting
\begin{equation}
\label{16}
v_t = \frac{\p}{\p x}\,\Bigl(f(t,x,v,v_1,\ldots,v_{n-1})\Bigr).
\end{equation}
So we arrive at the equation written in the form of conservation
law. Let us emphasize that the reason for an initial equation
(\ref{1}) to be transformable to the form (\ref{16}) is the contact
symmetry group (\ref{14}) admitted by (\ref{1}).

Now, provided evolution equation (\ref{15}) admits an additional
contact transformation group ${\mathfrak G}$
$$
\begin{array}{l}
t' = T(t,\theta),\quad x' = X(t,x,u,u_x,\theta),\\[2mm]
u' = U(t,x,u,u_x,\theta),\quad u'_{x'} = V(t,x,u,u_x,\theta),
\end{array}
$$
the latter is mapped into the symmetry group of evolution equation
(\ref{16}) by the non-local change of the dependent variable $v=\p
u/\p x$. The type of symmetry, Lie or non-Lie, is determined by the
coefficients $T,X,U,V$.

If
$$
\frac{\p T}{\p u} = \frac{\p X}{\p u} = \frac{\p U}{\p u} = \frac{\p
V}{\p u} = 0,
$$
then the group ${\mathfrak G}$ is mapped into the Lie symmetry
$$
t' = T(t,x,v,\theta),\quad
x' = X(t,x,v,\theta),\quad
v' = V(t,x,v,\theta)
$$
of equation (\ref{15}). Next, provided
$$
\left(\frac{\p T}{\p u}\right)^2 + \left(\frac{\p X}{\p u}\right)^2 +
\left(\frac{\p U}{\p u}\right)^2 + \left(\frac{\p V}{\p u}\right)^2 \ne 0,
$$
${\mathfrak G}$ is transformed into the non-local (potential)
symmetry
$$
t' = T(t,x,\p_x^{-1}v,v,\theta),\quad x' =
X(t,x,\p_x^{-1}v,v,\theta),\quad v' = V(t,x,\p_x^{-1}v,v,\theta).
$$
Hereafter, $\p_x^{-1}$ is the inverse of $\p_x$, so that
$\p_x\p_x^{-1} \equiv \p_x^{-1}\p_x\equiv 1$.

Rewriting the above listed constraints through infinitesimals of the
group ${\mathfrak G}$ we get the following assertion.

\begin{Theorem}
Let evolution equation (\ref{15}) be invariant under the group of
contact transformations
\begin{equation}
\label{17}
\begin{array}{l}
t' = t + \e\tau(t)\quad x' = x +\e\xi(t,x,u,u_x),\\[2mm]
u' = u + \e\eta(t,x,u,u_x),\quad u'_{x'} = u_x + \e\rho(t,x,u,u_x).
\end{array}
\end{equation}
Then differentiating (\ref{15}) with respect to $x$ and making the
non-local change of variables $v={\p u}/{\p x}$ yield an evolution
equation in a form of conservation law (\ref{16}), while group
(\ref{17}) is mapped into
\begin{itemize}
\item Lie symmetry of equation (\ref{16}), provided
$$
\frac{\p \tau}{\p u} = \frac{\p \xi}{\p u} = \frac{\p \eta}{\p u} =
\frac{\p \rho}{\p u} = 0;
$$
\item non-local (potential) symmetry of equation
(\ref{16}), provided
$$
\left(\frac{\p \tau}{\p u}\right)^2 + \left(\frac{\p \xi}{\p
u}\right)^2 + \left(\frac{\p \eta}{\p u}\right)^2 + \left(\frac{\p
\rho}{\p u}\right)^2 \ne 0.
$$
\end{itemize}
\end{Theorem}

It immediately follows from Theorem 1 that the problem of
constructing of all possible evolution equations that admit
potential symmetries is equivalent to classification of equations of
the form (\ref{16}) invariant under groups of contact
transformations
\begin{equation}
\label{18}
\begin{array}{l}
t' = t + \e\tau(t,x,u,u_x)\quad
x' = x +\e\xi(t,x,u,u_x),\\[2mm]
u' = u + \e\eta(t,x,u,u_x), \quad u'_{x'} = u_x + \e\rho(t,x,u,u_x)
\end{array}
\end{equation}
with
\begin{equation}
\label{20}
\left(\frac{\p \tau}{\p u}\right)^2 + \left(\frac{\p \xi}{\p
u}\right)^2 + \left(\frac{\p \eta}{\p u}\right)^2 + \left(\frac{\p
\rho}{\p u}\right)^2 \ne 0.
\end{equation}
Indeed, Theorem 1 states that any evolution type equation (\ref{16})
that is presented in the form of conservation law can be reduced to
the form (\ref{15}). What is more, its potential symmetry transforms
into contact symmetry (\ref{18}) of evolution equation (\ref{15}).
Based on this observation is our algorithm for group classification
of evolution equations admitting potential symmetries.

\begin{enumerate}
\item We compute the maximal symmetry group ${\mathfrak G}$ of
contact transformations leaving differential equation (\ref{1})
invariant.
\item We classify inequivalent one-parameter subgroups of
${\mathfrak S}$ and select subgroups ${\mathfrak
G}_1,\ldots{\mathfrak G}_p$ of the form (\ref{14}).
\item For each subgroup, ${\mathfrak G}_i$, we construct change of
variables (\ref{19}) reducing the corresponding subgroup to the
group of translations by $u$, which leads to evolution equations of
the form (\ref{16}).
\item Since the invariance group, $\bar{\mathfrak G}$, admitted by
(\ref{15}) is isomorphic to ${\mathfrak G}$, we can utilize the
results of subgroup classification of ${\mathfrak G}$. For each of
the one-parameter subgroups of $\bar{\mathfrak G}$ we check whether
its infinitesimals (\ref{18}) satisfy (\ref{20}). This yields the
list of evolution equations that can be reduced to those admitting
potential symmetries.
\item Performing the non-local change of variables $v={\p u}/{\p x}$
yields evolution equations (\ref{16}) admitting potential
symmetries.
\end{enumerate}

The above algorithm takes especially simple form for the case when
the maximal group, ${\mathfrak G}$, admitted by evolution equation
(\ref{1}) contains Lie symmetries only. Suppose that evolution
equation (\ref{1}) admits at least a two-parameter Lie symmetry group
${\mathfrak G}$. Furthermore, we suppose that this group has a
one-parameter subgroup leaving the variable $t$ invariant. With this
condition in hand we can reduce equation under study to the form
(\ref{15}). The transformed equation (\ref{15}) admits at least two
Lie symmetries $e_1=\p_u$ and $e_2 = \tau(t)\p_t + \xi(t,x,u)\p_x +
\eta(t,x,u)\p_u$. Now, if the infinitesimals $\tau, \xi, \eta$
satisfy one of the inequalities
\begin{eqnarray}
&&\frac{\partial \xi}{\partial u} \not=0, \label{21}\\
&&\frac{\partial\xi}{\partial u}=0,\quad
\left(\frac{\partial^2\eta}{\partial u\partial x}\right)^2 +
\left(\frac{\partial^2\eta}{\partial u\partial u}\right)^2 \not=
0,\label{22}
\end{eqnarray}
then (\ref{15}) reduces to the conservation law form (\ref{16}), and
what is more, the Lie symmetry $e_2$ is mapped into the potential
symmetry of the obtained evolution equation (see, also \cite{ren3}).

\section{Examples.}

As shown in \cite{ren1} the maximal Lie symmetry algebra of the
second-order evolution equation
\begin{equation}
\label{23} u_t = u_{xx}-uu_x+\lambda u_x^{3/2},\quad \lambda\in
{\mathbb R},\quad \lambda\ne 0
\end{equation}
is spanned by the following infinitesimal operators
\begin{equation}
\label{24}
e_1=\p_t,\quad e_2=\p_x,\quad e_3=t\p_x+\p_u,\quad
e_4 = 2t\p_t+x\p_x-u\p_u.
\end{equation}
This algebra has two inequivalent one-dimensional subalgebras that
preserve the variable $t$, namely, $A_1=\langle \p_x\rangle$ and
$A_2=\langle t\p_x+\p_u\rangle$. Consider, first, the algebra $A_1$.
As a first step of the algorithm we need to transform the basis
operator of $A_1$ to the canonic form. This is achieved by the
following change of variables
$$
\bar t = t,\quad \bar x = u,\quad \bar u = x.
$$
Now algebra (\ref{24}) reads as
$$
e_1=\p_t,\quad e_2=\p_u,\quad e_3=t\p_u+\p_x,\quad
e_4 = 2t\p_t+u\p_u-x\p_x
$$
(we dropped the bars). Evidently, the coefficients of the operators
$e_1, e_3, e_4$ do not obey the conditions (\ref{21}), (\ref{22}),
which means that the operator $e_2=\p_x$ does not yield an equation
possessing potential symmetries.

Turn now to the second algebra $A_2$. The change of variables
\begin{equation}
\label{25}
\bar t = t,\quad \bar x = x-tu,\quad \bar u = u
\end{equation}
reduces its basis element $t\p_x+\p_u$ to the canonic form $\p_{\bar
u}$. After making in (\ref{24}) the above change of variables, we
get
$$
e_1=\p_t-u\p_x,\quad e_2=\p_x,\quad e_3=\p_u,\quad
e_4 = 2t\p_t+u\p_u-x\p_x
$$
(as earlier, we drop the bars). Now the coefficients of the operator
$e_1$ satisfy (\ref{21}) and, consequently, change of variables
(\ref{25}) transforms equation (\ref{23}) to the one admitting the
potential symmetry.

Indeed, being written in the new variables equation (\ref{23}) takes
the form
$$
u_t = (1+tu_x)^{-2}u_{xx} + \lambda (1+tu_x)^{-1/2}u_x^{3/2}.
$$
Differentiating this equation with respect to $x$ and introducing
the new dependant variable $v(t,x)=u_x$, we get
$$
v_t=\frac{\p}{\p x}\Bigl((1+tv)^{-2}v_{x} + \lambda
(1+tv)^{-1/2}v^{3/2}\Bigr).
$$
This is the evolution equation represented in the form of
conservation law and it admits the following one-parameter group of
non-local transformations
$$
t' = t+\theta,\quad x' = x-\theta \p_x^{-1}v,\quad v' = v.
$$
Here $\theta$ is the group parameter.

Consider now applying our algorithm to the following $sl(2, {\mathbb
R})$ invariant evolution equation \cite{ren4}:
\begin{equation}
\label{26}
u_t = xu_xf(t,\omega),\quad
\omega=x^{-5}u_x^{-3}u_{xx}+2x^{-6}u_x^{-2}.
\end{equation}
Under arbitrary $f$ the maximal symmetry algebra admitted by
equation (\ref{26}) reads as
$$
e_1=\p_u,\quad e_2=2u\p_u-x\p_x,\quad
e_3=-u^2\p_u+xu\p_x.
$$
It is a common knowledge that there exist three inequivalent
one-dimensional subalgebras of the algebra $sl(2,{\mathbb R})$,
namely, $\langle e_1\rangle,$ $\langle e_2\rangle,$ $\langle
e_1+e_3\rangle$.

Consider first the algebra $\langle e_1\rangle$. Its basis operator
is already in the canonic form. Since the coefficients of the basis
element $e_3$ satisfy (\ref{22}), it leads to the evolution equation
that admits a potential symmetry. The equation in question is
obtained if we differentiate (\ref{26}) with respect $x$ and
introduce the new dependant variable $v=u_x$
$$
v_t = \frac{\p}{\p x}\Bigl(xvf(t,\omega)\Bigr),\quad
\omega=x^{-5}v^{-3}v_{x}+2x^{-6}v^{-2}.
$$

Turn now to the algebra $e_2$. Making the change of variables
\begin{equation}
\label{27}
\bar t = t,\quad \bar x = x^2u,\quad \bar u = \frac{1}{2}\ln u
\end{equation}
reduces $e_2$ to the canonic form $\p_{\bar u}$. The operators $e_1$
and $e_3$ take the form
$$
e_1=x\exp(-2u)\p_{x}+\frac{1}{2}\exp(-2u)\p_{u},\quad e_2=-
x\exp(2u)\p_{x}+\frac{1}{2}\exp(2u)\p_{u}.
$$
Note that in the above formulas we drop the bars. Since the
coefficients of the operators $e_1, e_3$ satisfy constraints
(\ref{22}) they both lead to the potential symmetries of the
evolution equation obtained from (\ref{26}) by making the change of
variables (\ref{27}), then differentiating the obtained relation
with respect to $\bar x$ and, finally, replacing $u_x$ with $v$. As
a result, we get the differential equation
$$
v_t = \frac{\p}{\p x}\Bigl(2xvf(t,\omega)\Bigr),\quad \omega =
2x^{-3}v^{-3}v_x + 3x^{-4}v^{-2} - 4x^{-2},
$$
which admits the two-parameter group of potential symmetries.

Finally, consider the algebra $\langle e_1+e_3\rangle$. The change
of variables
\begin{equation}
\label{28} \bar t = t,\quad \bar x = x^2(1-u^2),\quad \bar u =
\frac{1}{2}\ln\frac{1-u}{1+u}
\end{equation}
reduces the operator $e_1+e_3$ to the canonic form $\p_{\bar u}$.
And what is more, the basis operators $e_1, e_2$ now read as
$$
e_1 = x\sinh 2u\p_{x}-\cosh^2 u\partial_{u},\quad e_3 = -2 x\cosh
2u\p_x + \sinh 2u\p_{u}.
$$
As usual, we drop the bars. Making change of variables (\ref{28}) in
the initial evolution equation, differentiating with respect to $x$
and replacing $u_x$ with $v$ yield
$$
v_t = \frac{\p}{\p x}\Bigl(-2xvf(t,\omega)\Bigr),\quad \omega =
2x^{-3}v^{-3}v_x+6x^{-4}v^{-2}-4x^{-2}.
$$
Since the coefficients of the operators $e_1$ and $e_3$ satisfy
constraints (\ref{22}), the above evolution equation admits the
two-parameter group of potential symmetries.

\section {Concluding Remarks}

Group approach to classification of evolution equations developed in
the present paper can easily be modified to become
applicable to systems of evolution equations in the same way as it
has been done for quasi-local symmetries in \cite{ren3}. The
essential difference is that for the case of systems of evolution
equations the potential symmetries always correspond to Lie
symmetries of transformed system of evolution equation. The reason
is that if a system of partial differential equations admits a group of
contact symmetries, the latter is always a first prolongation of a
Lie symmetry group \cite{ibr1}. Hence it follows, in particular,
that the problem of classification of systems of evolution type
equations admitting potential symmetries can be solved completely
within the framework of point Lie symmetries.

In the case of a single evolution equation we should go beyond Lie
symmetry in order to recover all possible equations admitting
potential symmetries. One needs to classify inequivalent
finite-dimensional groups of contact transformations realized as
symmetry groups of (\ref{1}). While the problem of group
classification of equations (\ref{1}) that admit Lie symmetry is
well understood and there are powerful methods to handle it, 
systematic exploration of contact symmetries of evolution equations
is yet to be done. Some partial results on this topic can be found
in \cite{sok1}-\cite{mon1}.

We intend to devote one of our future publication to application of
the approach developed in the present paper to utilize contact
symmetries of evolution equations in order to classify their
non-local symmetries.

\end{document}